%% file: main.tex
\pdfoutput=1
\documentclass{article}
\usepackage{spconf,amsmath,graphicx}
\usepackage{siunitx}
\usepackage{tabularx}
\usepackage{pifont}
\usepackage[skip=3pt]{caption}
\usepackage{booktabs}
\usepackage{xcolor}
\usepackage{hyperref}
\usepackage{cleveref}
\usepackage{tikz}
\usepackage{pgfplots}
\usepackage{bm}
\usepackage{upgreek}
\usepackage{etoolbox}
\usepackage{multirow} 
\usetikzlibrary{positioning, fit, calc}
\usetikzlibrary{patterns}
\usepackage[acronym, shortcuts, toc,nonumberlist]{glossaries}

\glsdisablehyper    
\newacronym{PLDA}{PLDA}{Probabilistic Linear Discriminant Analysis}
\newacronym{SC}{SC}{Spectral Clustering}
\newacronym{VMFMM}{VMFMM}{von-Mises-Fischer Mixture Model}
\newacronym{VAD}{VAD}{Voice Activity Detection}
\newacronym{DER}{DER}{Diarization Error Rate}
\newacronym{EER}{EER}{Equal Error Rate}
\newacronym{EM}{EM}{Expectation-Maximization}
\newacronym{TAP}{TAP}{Time Average Pooling}
\title{Geodesic interpolation of frame-wise speaker embeddings \\for the diarization of meeting scenarios}
\name{%
\begin{tabular}{@{}c@{}}
Tobias Cord-Landwehr$^{\star}$, %
Christoph Boeddeker$^{\star}$, %
C\u{a}t\u{a}lin Zoril\u{a}$^{\dagger}$, %
Rama Doddipatla$^{\dagger}$,\\%
Reinhold Haeb-Umbach$^{\star}$%
\end{tabular}
}
\address{${}^{\star}$ Paderborn University, Department of Communications Engineering, Paderborn, Germany\\
${}^{\dagger}$ Toshiba Europe Ltd, Cambridge, United Kingdom}

\begin{document}

\ninept
\abovedisplayskip7.5pt plus 3.0pt minus 4.0pt
\belowdisplayskip7.5pt plus 3.0pt minus 4.0pt

\setlength{\floatsep}{1ex}
\setlength{\textfloatsep}{0.9em}

\setlength{\abovedisplayskip}{3pt}
\setlength{\belowdisplayskip}{3pt}
\setlength{\textfloatsep}{9pt plus 0.0pt minus 2.0pt}
\setlength{\floatsep}{7pt plus 0.0pt minus 2.0pt}
\setlength{\intextsep}{10pt plus 0.0pt minus 2.0pt}

\abovedisplayskip7.5pt plus 3.0pt minus 4.0pt
\belowdisplayskip7.5pt plus 3.0pt minus 4.0pt

\makeatletter
\renewcommand\section{\@startsection {section}{1}{\z@}%
                                   {-3.2ex \@plus -1ex \@minus -1.5ex}%
                                   {1.9ex \@plus.0ex}%
                                   {\normalfont\Large\bfseries}}
\renewcommand\subsection{\@startsection{subsection}{2}{\z@}%
                                     {-2.ex\@plus -1ex \@minus -.2ex}%
                                     {1.ex \@plus .2ex}%
                                     {\normalfont\large\bfseries}}
\makeatother
\maketitle
\begin{abstract}
We propose a modified teacher-student training for the extraction of frame-wise speaker embeddings that allows for an effective diarization of meeting scenarios containing partially overlapping speech.
To this end, a geodesic distance loss is used that enforces the embeddings computed from regions with two active speakers to lie on the shortest path on a sphere between the points given by the d-vectors of each of the active speakers.
Using those frame-wise speaker embeddings in clustering-based diarization outperforms segment-level clustering-based diarization systems such as VBx and Spectral Clustering.
By extending our approach to a mixture-model-based diarization, the performance can be further improved, approaching the diarization error rates of diarization systems that use a dedicated overlap detection, and outperforming these systems when also employing an additional overlap detection.

\end{abstract}
\section{Introduction}
\label{sec:intro}
Current approaches for the diarization of meeting scenarios can be categorized into either frame-level neural network-based, also called End-to-End, diarization systems \cite{19_Fujita_EEND, 23_chen_attention_eend}, or segment-level clustering-based \cite{22_landini_vbx, 23_park_voxrsc23_gist} diarization systems.

The former systems natively output frame-level diarization estimates and are able to assign multiple active speakers to each time frame.
However, they suffer from the limitation that matching training data is necessary so they generalize poorly to different domains. Also, they are prone to represent speakers only on a local, not a global level, which complicates the re-identification of speakers over long durations, so current works try to address this bottleneck \cite{eda_eend_local_global}. 

The latter models extract a single embedding, such as x-vectors or d-vectors \cite{18_snyder_xvector, 17_li_deep_embeddings}, for each audio segment of a meeting recording. Then, clustering is employed to assign a speaker label to each segment. 
Through the speaker embedding extractor, which is typically trained out of domain on a large number of speakers, these systems generalize well to mismatched conditions and are able to maintain a global speaker representation. 
However, these systems are less suited to model fast speaker turns or overlapping speech due to the segment-level embedding extraction.
While earlier diarization pipelines used a simple \gls{VAD} followed by a uniform segmentation \cite{17_garcia_xvec_diarization},
current systems address this problem by using neural network-based \cite{21_bredin_segmentation} or multi-scale segmentation models \cite{22_kwon_multiscale_segmentation} to obtain fine-grained activity information and map the speaker labels accurately to the active speech regions.
Though there are also approaches to include information about overlapping speech in this step, it is still considered as postprocessing to refine the single-speaker labels. Therefore, especially back channels, where another speaker briefly interrupts the active speaker, are hard to handle with these approaches.
Some current systems also incorporate both approaches in order to evade these drawbacks \cite{21_bredin_segmentation, 22_kinoshita_eend_vc}.
While in \cite{23_cordlandwehr_framewise_embeddings} it was demonstrated that frame-wise embeddings bring advantages to End-to-End diarization, here, we consider the use of frame-wise embeddings in clustering-based processing. 

In this work, we aim to forego the segmentation model of such a pipeline and directly perform a clustering-based diarization using frame-level speaker embeddings obtained from a teacher-student training.
To this end, we propose a new training objective to regularize the behavior of the frame-wise speaker embeddings in the presence of overlapping speech.
Through this loss, the frame-wise speaker embeddings are encouraged to lie on the geodesic, i.e., the path along the hypersphere, between single-speaker embeddings of the two active speakers.
Then, a simple k-Means clustering is used to 
provide a diarization directly based on the frame-level embeddings. Through its hard-decision class assignment, k-Means clustering is unable to properly address overlapping speech. We thus move to a mixture model-based clustering to better account for overlapping speech regions. 
Here, we model each speaker by a von-Mises-Fischer distribution and use the class posterior probabilities as soft labels for a diarization system. Since the performance of mixture models is known to be sensitive to their initialization \cite{baudry2015mixtures}, we then use the k-Means clusters for initializing the distribution parameters of the mixture model to obtain an overall more robust system.

This work is structured as follows. First, \Cref{sec:embeddings} describes the teacher-student training for the extraction of frame-wise embeddings as well as its extension for the training on overlapping speech segments.
Then, \Cref{sec:diarization} covers the diarization using these frame-wise speaker embeddings.
In \Cref{ssec:eval}, the proposed frame-level diarization pipeline and the segment-level reference systems are explained. Then, the proposed model is evaluated on the LibriCSS and DipCo datasets.
Finally, \Cref{sec:conclusion} gives a summary and outlines future research perspectives.

\section{Frame-wise speaker embedding extraction}
\label{sec:embeddings}
The teacher-student approach from \cite{23_cordlandwehr_framewise_embeddings} is used to extract frame-level speaker embeddings from meeting data. 
First, a ResNet34-based d-vector teacher \cite{17_li_deep_embeddings} is trained using an AAM-Softmax \cite{liu2019AAM} classification loss.
Then, the weights of the teacher are kept frozen, and the frame-wise student model is trained using speaker embeddings provided by the teacher model.
The student has the same architecture as the teacher, except that the \gls{TAP} at the model's output is exchanged with a local pooling that averages over a few neighboring time frames (here: \num{11}) while maintaining a frame advance of one frame.
Given the classification loss employed during training, the length-normalized teacher embeddings lie on a hypersphere.
Therefore, the goal of the student is to project each frame-wise speaker embedding into this latent space so that the frame-wise embeddings also can be assumed to share the same geometrical properties as the teacher embeddings.

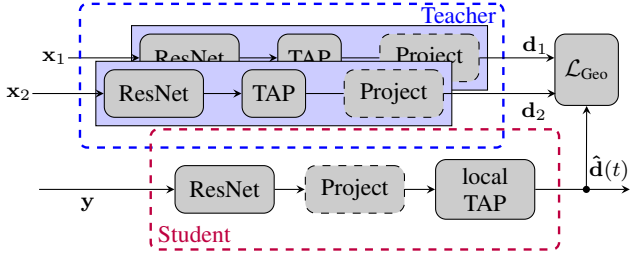
\begin{figure}[bt]
\input{tikz/teacher_student}
	\caption{Visualization of the modified Teacher-Student training for partially overlapping training data}
  \label{fig:teacher_student}
\end{figure}

\subsection{Geodesic Speaker Loss}
In \cite{23_cordlandwehr_framewise_embeddings}, it was observed that, even though 
the student is only trained on single speaker regions, the frame-wise embeddings smoothly transfer from one to another speaker in the presence of overlapping speech. However, there was no control over how exactly this occurred. 
In order to explicitly encourage the frame-wise embedding to reflect the two competing speakers, in this work, the student model is directly trained on partially overlapping speech. The MSE-based loss $\mathcal{L}_{\mathrm{MSE}}$ used in \cite{23_cordlandwehr_framewise_embeddings} is replaced by a geodesic loss function as follows:
The loss specifies the closest speaker embedding $\mathbf{d}(\tilde{\alpha}_t)$ as target embedding for overlap,
where the target is chosen as 
\begin{align}
      \mathbf{d}(\tilde{\alpha}_t) &= \tilde{\alpha}_t \mathbf{d}_1 + (1-\tilde{\alpha}_t) \mathbf{d}_2; \quad \text{where}\\
    \tilde{\alpha}_t &= \operatornamewithlimits{argmin}_{\alpha_t \in [0,1]} \lVert\mathbf{d}(\alpha_t)- \hat{\mathbf{d}}_t\rVert^2.    
\end{align}
Thus, $\mathbf{d}(\tilde{\alpha}_t)$ is the linear interpolation between the embeddings $\mathbf{d}_1$ and $\mathbf{d}_2$ of the speakers active in the overlap region, that has the smallest Euclidean distance to the current estimate of the frame-wise embedding 
$\hat{\mathbf{d}}_t$.
The optimal $\tilde\alpha_t$ is obtained per frame through a constrained least squares problem. 
This target is then scaled back to the length of the teacher embeddings by multiplication with the scalar $r$: 
\begin{align}
    r = \tfrac{\lVert \mathbf d_1\rVert + \lVert\mathbf d_2\rVert}{2}.
\end{align}
Through this loss, the student embeddings are encouraged to lie on the geodesic, i.e., the path on the hypersphere between both active speakers, for overlapping speech.
For single-speaker regions, only the MSE-loss is used. Overall, the proposed training objective is thus given by:
\begin{align}
\mathcal{L}_{\mathrm{geo}} = \sum_t
\begin{cases}
    \lVert r\frac{\mathbf{d}(\tilde{\alpha}_t)}{\lVert \mathbf{d}(\tilde{\alpha}_t)\rVert} - \hat{\mathbf{d}}_t\rVert^2 \quad & \text{if overlap}, \\
    \lVert \mathbf d_1 - \hat{\mathbf{d}}_t\rVert ^2 \quad & \text{if spk1}, \\
    \lVert \mathbf d_2 - \hat{\mathbf{d}}_t\rVert ^2 \quad & \text{if spk2.}
\end{cases} 
\end{align}


\subsection{Dimension reduction layer}
For clustering-based diarization systems, it is a common practice to project the speaker embeddings into a lower dimensional space to obtain more robust embeddings for clustering \cite{22_landini_vbx, 23_park_voxrsc23_gist, 17_garcia_xvec_diarization}. 
Since the main idea behind the frame-wise embeddings is to exploit their latent structure and geometry along a hypersphere, popular dimension reduction techniques such as \gls{PLDA} cannot be employed since they do not maintain the latent geometry of the data. 
Instead, an additional projection layer is added after the teacher to reduce the \num{256}-dimensional speaker embeddings down to \num{64} dimensions. This projection layer is fine-tuned with an AAM-Softmax loss while keeping the remainder of the teacher weights fixed. Then, the student model is trained to reproduce the lower-dimensional speaker embeddings by incorporating an additional projection layer at the student ResNet's output.  An illustration of the modified teacher-student training is depicted in \Cref{fig:teacher_student}.

\section{Diarization of Frame-level embeddings}
\label{sec:diarization}
Compared to other diarization pipelines, we forego segmentation and directly cluster the frame-wise speaker embeddings to obtain the diarization result. Therefore, no additional steps to map a segment-level assignment back to the diarization estimate are necessary.   
In the simplest approach, overlap is assumed to be absent and each frame is assigned to only one active speaker.
In this case, a simple k-Means algorithm can be used to directly solve the diarization objective.
First, a VAD is applied to only retain the frame-wise embeddings that contain speech.
Then, k-Means++ \cite{07_arthur_kmeans} is applied to the active embeddings to obtain a class label for each embedding.
These labels are then used as hard decisions and directly provide the diarization estimate.
Therefore, this approach inherently ignores overlapping speech since each frame is only assigned to a single speaker.

\subsection{Mixture model-based clustering}
A mixture model-based approach can be seen as the soft decision generalization of k-Means.
Instead of assigning labels to each embedding in order to minimize the total distance to the cluster centers, the observations are modeled as draws from a mixture of probability distributions.
Here, the class labels are assigned by their likelihood of belonging to a mixture component.  
The von-Mises-Fischer distribution  
\begin{align}
	p(\bar{\mathbf d}_t; \boldsymbol{\upmu}_k, \kappa_k) &= c_E(\kappa_k)e^{\kappa_k\boldsymbol{\upmu}_k^\mathrm{T}\bar{\mathbf{d}}_t} \\
	c_E(\kappa_k) &= \frac{\kappa_k^{E/2 -1}}{(2\pi)^{E/2}I_{E/2}(\kappa_k)} 
\end{align}
with the regular, first-order Bessel function $I_n(\cdot)$
describes a probability distribution on the $E$-dimensional hypersphere. Here, $\boldsymbol{\upmu}_k$ specifies the main direction of the distribution, and the concentration parameter $\kappa$ determines its spread across the hypersphere.
Therefore, it poses a natural choice as a distribution 
for diarization, since the
length-normalized student embeddings $\bar{\mathbf{d}}_t$ of a single speaker $k$ can be assumed as realizations of such a distribution with $\boldsymbol{\upmu}_k$ describing the average embedding and $\kappa_k$ accounting for the spread of the speaker embeddings.
Using this assumption, the likelihood of all frame-wise embeddings in a meeting scenario can be modeled with the mixture of von-Mises-Fischer distributions
\begin{align}
	\ell(\bar{\mathbf{d}};\boldsymbol{\upmu}, \boldsymbol{\kappa}) = \prod_t\sum_{k=1}^K\pi_k p(\bar{\mathbf{d}}_t; \boldsymbol{\upmu}_k, \kappa_k)
\end{align}
 for a meeting with $K$ speakers.
Such a \gls{VMFMM} can be fitted to the observations using the \gls{EM} algorithm \cite{05_banerjee_vmf}, which alternates between calculating the class posterior probabilities $\gamma_{tk}$, also called class affiliations, given the current estimates of the mixture model parameters,  and updating the model parameters given the updated class posterior probabilities \cite{McLachlan1996TheEA}.
After convergence, the class affiliations $\gamma_{tk}$ can be interpreted as soft labels indicating the probability of a given frame $t$ belonging to class $k$.
Since the geodesic loss encourages speaker embeddings to lie between both active speakers in regions of overlapping speech, it can be assumed that the class affiliations for both active speakers have a contribution and thresholding can be applied to the posterior probabilities to assign multiple active speakers to these frames.

\section{Experiments}
\label{ssec:eval}

\subsection{Evaluation Setup}
The teacher model is trained on VoxCeleb \cite{Nagrani19} segments of \SI{4}{\second} that are augmented with MUSAN \cite{snyder2015musan} noise and simulated room impulse responses \cite{habets2006room}. The 256-dimensional d-vector system achieves an \gls{EER} of \SI{1.1}{\percent} on the VoxCeleb1-O evaluation set, while using the projection layer to reduce the dimensionality to \num{64} slightly reduces the performance to \SI{1.5}{\percent} \gls{EER}.
Then, the frame-wise embedding extractor is trained on partially overlapping VoxCeleb two-speaker mixtures with an overlap of \SIrange{20}{40}{\percent} using the teacher embeddings as a target.
For the evaluation of the diarization using the frame-wise embeddings, a simple diarization pipeline is constructed.
First, all frame-wise embeddings are extracted from a meeting recording and are normalized to zero mean and unit length across the time axis.
Then, an energy-based \gls{VAD} using minimum statistics is employed on the meeting audio to discard silence regions from the meeting data. The threshold of this \gls{VAD} was chosen such that it consistently underestimates the activity since the embedding extractor is not trained on silence.
Afterward, the frame-level embeddings containing speech are used for clustering. Here, the total number of speakers in the meeting is assumed to be known.
For k-Means clustering, the class labels are directly mapped to the diarization estimates. For the mixture model-based diarization, the class affiliations are used as diarization estimates after thresholding at \num{0.3}.
Then, a maximum filter of \SI{1.3}{\second} followed by a minimum filter of \SI{1}{\second} is applied to the diarization estimates to fill gaps in the estimated activity and revert effects of the underestimated \gls{VAD}. 

This pipeline is evaluated both on the LibriCSS database \cite{20_Chen_libricss} containing simulated meetings of LibriSpeech re-recordings with different overlap ratio subsets, and the DipCo \cite{19_segbroeck_dipco} dinner party meeting corpus that was used in the 7th CHiME challenge \cite{23_cornell_chime7}. For LibriCSS, the first session of all subsets is used as a development set to optimize the model hyperparameters.
As an evaluation metric, the \gls{DER} is calculated according to \cite{21_nist_sre_plan}. For LibriCSS, no collar is used, and for DipCo, a collar of \SI{250}{\milli\second} is chosen as in \cite{23_cornell_chime7}.

\subsection{Reference Systems}
As reference systems, the proposed frame-wise diarization is compared to segment-level clustering-based diarization systems with several different clustering approaches.
Here, both \gls{SC} and VBx \cite{19_park_sc, 22_landini_vbx}, as well as their extensions with a dedicated overlap detection proposed in \cite{21_raj_scovl, 20_bullock_vbxovl} are used as reference systems.
The \gls{SC} models are set to \num{8} speakers, while VBx infers the number of speakers automatically.
All reference models use the same general processing pipeline.
First, a neural network-based segmentation model \cite{21_bredin_segmentation} is applied to the meeting data to determine speech regions and regions containing overlapping speech.
Then, a ResNet101 trained on VoxCeleb is used to extract d-vectors for each segment containing speech.
Afterward, a \gls{PLDA} is used to reduce the dimensionality of the embeddings, followed by a clustering (i.e.\ \gls{SC} or VBx) to obtain the labels of each segment. Finally, the activity obtained by the segmentation model is used to map the class labels back to a diarization estimate. With overlap detection, the estimated overlap regions are additionally used during the clustering stage to assign two speaker labels to these regions.

\begin{table}[bt]
    \centering
        \caption{Influence of the geodesic loss $\mathcal{L_{\mathrm{geo}}}$ on the diarization error rate (DER) on LibriCSS. 
        }
    \label{tab:geodesic_loss}
    \begin{tabular}{l l S S}
    \toprule
        Model & Loss &  \multicolumn{2}{c}{DER [\si{\percent}]} \\
        \cmidrule{3-4}
        & & {dev} & {test} \\
    \midrule
         k-Means & $\mathcal{L}_{\mathrm{MSE}}$ & 90.78 & 87.59\\
         k-Means & $\mathcal{L}_{\mathrm{geo}}$ & 20.05 & 16.53\\
     \midrule
         VMFMM   & $\mathcal{L}_{\mathrm{MSE}}$ & 98.82 & 86.83 \\
         VMFMM   & $\mathcal{L}_{\mathrm{geo}}$ & 24.07 & 25.23 \\ 

    \bottomrule
    \end{tabular}

\end{table}

\subsection{Effect of the geodesic loss}

First, the influence of using the geodesic loss is investigated. \Cref{tab:geodesic_loss} shows the comparison of using the frame-wise student embeddings from \cite{23_cordlandwehr_framewise_embeddings} to the speaker embeddings using the geodesic training loss for clustering, instead. It becomes apparent that the geodesic loss is instrumental in order to extract meaningful cluster labels from the frame-wise embeddings.  

\begin{table*}[hbt]
	\caption{Comparison of the DER of segment-level speaker diarization systems (rows 1 -- 4)\protect\footnotemark[1] against the frame-wise diarization (rows 5 -- 11) on the first channel of the LibriCSS datasets.
\textit{avg.}, \textit{single}, and \textit{OV} denote the \gls{DER}, the  \gls{DER} on single-speaker regions, and the \gls{DER} considering only overlapping speech, averaged over all sessions, respectively. The \gls{VMFMM} uses  $\kappa_{\mathrm{max}}=\num{25}$ and \num{50} EM iterations.}
 \label{tab:mm_results}
\centering
    \robustify\bfseries
    \sisetup{detect-weight=true, detect-family=true}
 \begin{tabular}{c l S S S S S S | S S S}
 \toprule
     row \# &Model & {0S} & {0L}& {OV10} & {OV20} & {OV30} & {OV40} & {avg.} & {single} & {OV}\\
     \midrule
     1 &SC \cite{19_park_sc} &  6.22 & 6.87 & 11.88 & 15.10 & 19.30 & 24.36 & 14.86 & 6.27 & 51.80\\
     2 & + ov detection \cite{21_raj_scovl}  & 6.28 & 7.09 & 10.73 & 11.86 & 13.22 & 15.74 & 11.28 & 7.34 & 28.20\\
     3 & Vbx\footnotemark[2] \cite{22_landini_vbx} & 6.39 & 5.94 & 10.42 & 15.07 & 19.29  & 23.99 & 14.45 & 5.90 & 49.69 \\ 
     4 & + ov detection\footnotemark[2] \cite{20_bullock_vbxovl}& 6.71 & 6.34 & 9.31 & 11.39 & 14.08 & 16.13 & 11.18 & 7.11 & 28.70 \\
       \midrule
      5 & k-Means (E=256) & 9.36 & 9.23 & 12.39 & 17.31 & 22.11 & 21.93& 16.53& 9.67 & 46.10\\
      6 & $\rightarrow$ E=64 &  4.62 & 7.27 & 10.25 & 13.84& 16.26 &21.08 & 12.98 & 5.23 & 46.39 \\
     \midrule
     7 & VMFMM (E=256) & 19.11 & 20.54& 24.86 & 27.53 & 25.92 & 31.03 & 25.23 & 20.05 & 47.52 \\
      8 & $\rightarrow$ E=64 & 16.58 & 17.27 & 11.20 & 23.06 & 22.66 & 26.43 & 19.89 & 15.88 & 49.59\\
      9 & + overinit  & 8.83 & 10.41 & 13.43 & 17.72 & 19.69 &  23.70 & 16.15 & 9.10 & 46.53\\
      10 & + k-Means init & 5.17 & 6.06 & 8.68  & 14.57 & 13.85 & 15.52 &11.07 & 6.17 & 32.18 \\ 
      11 & $\quad$+ ov detection & 4.75 &  6.92& 8.61 &  11.01 &   13.39&  15.21 &  10.36 & 5.41 & 31.84 \\ 
    \bottomrule
 \end{tabular}
\end{table*}

\subsection{Comparison to segment-level diarization techniques}
\Cref{tab:mm_results} depicts the comparison of the proposed frame-wise clustering to the typically used segment-level approaches.
It can be seen that a simple k-Means clustering of the frame-wise embeddings already provides a solid diarization (row 5). 
However, the 256-dimensional embeddings prove to be unable to be directly used for a robust diarization. By using the projected, \num{64}-dimensional speaker embeddings  (row 6), the diarization performance is significantly improved and a k-Means clustering of the frame-wise speaker embeddings outperforms both \gls{SC} and VBx.
The \gls{VMFMM}, however, achieves much worse results compared to a k-Means clustering (row 7). Again, using lower-dimensional embeddings improves the performance (row 8). 
When directly comparing the k-Means clustering and VMFMM, the performance of the mixture model is adversely affected by outliers, where one speaker is split into two mixture components and another speaker is ignored entirely. To alleviate this issue, the \gls{VMFMM} is initialized with one additional component, and the two mixture components with the highest intersection-over-union ratio are fused after \num{20} iterations (overinit, row 9). This significantly reduces the number of outliers in the results and improves the diarization performance. However, the randomly initialized mixture model still performs worse than k-Means clustering.

\subsection{k-Means initialization of the mixture model}
Mixture models are known to be sensitive to their initialization. In case of an unfavorable initialization, they are known to converge to a poor local minimum.
First, k-Means clustering is employed on the frame-wise embeddings. Then, the cluster centers of k-Means are used to initialize the $\boldsymbol{\upmu}_k$ of the \gls{VMFMM} with a $\kappa_k$ of \num{10}, each. 
Using this initialization results in an additional improvement upon only using k-Means, outperforming the segment-level diarization systems that use a dedicated overlap detection (row 10).
Here, the performance gains stem both from single-speaker regions and regions of overlapping speech. Therefore, choosing an appropriate initialization for the mixture model not only helps to obtain more robust single-speaker representatives but also results in overlapping mixture components from which overlap can be inferred.
The overlap diarization error still is worse than for systems with an overlap detection. Therefore, the external overlap detection from \cite{20_bullock_vbxovl} can be used to identify overlap regions and explicitly set the two most likely classes of these regions as active.   
In doing so, the \gls{VMFMM} again is slightly improved, furthering the gap both to the overlap-aware \gls{SC} and VBx system (row 11).

\footnotetext[1]{Implementation taken from https://github.com/desh2608/diarizer. 
}
\footnotetext[2]{Number of speakers estimated by clustering}

\subsection{Impact of the concentration parameter}
\begin{figure}[bt]
    \input{tikz/concentration_performance.tex}
	\caption{DER of the VMFMM on LibriCSS for different maximal $\kappa$ 
 }
    \label{fig:concentration}
\end{figure}
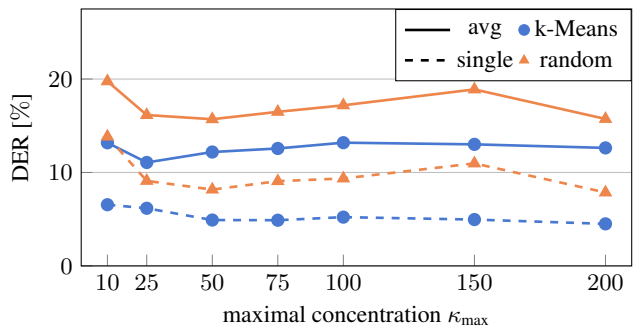

For the parametrization of the mixture model, the concentration parameter $\kappa$ was limited to a value $\kappa_{\mathrm{max}}=25$ for all mixture components. Without this limit, the concentration of single components grows too large, resulting in a worse overlap \gls{DER} and an overall worse performance.
This indicates a trade-off of the mixture model. On the one hand, the mixture components need to be able to accurately portray the embedding spread for a speaker. Therefore, limiting the extent of the concentration parameter too much results in less descriptive mixture components and a higher \gls{DER}. \Cref{fig:concentration} illustrates this behavior. It also shows that the \gls{DER} stagnates or even degrades for high values of $\kappa_{\mathrm{max}}$. In this case, the mixture components become sharper and intersect less with each other, which can improve the single-speaker performance. At the same time, the class affiliations converge to hard decisions and overlap can no longer be inferred from them.
So, in order to optimize the performance of the mixture model, a trade-off between single-speaker and overlap performance needs to be found.
Initializing the \gls{VMFMM} through k-Means makes this effect more visible since the model is less sensitive to outliers. 


\subsection{Comparison on DipCo}
Finally, the frame-wise clustering model is applied to the DipCo data. Here, only the first of \num{35} microphone channels is used for clustering, and the CHiME7 baseline \gls{VAD} is used.
\Cref{tab:dipco_results} shows that a simple k-Means clustering proves much worse on this dataset, which is expected since the data contains a higher amount of overlapping speech. Therefore, the \gls{VMFMM} does not profit from an initialization with k-Means. Still, the mixture model outperforms the CHiME7 diarization baseline by more than \SI{5}{\percent} and \SI{2}{\percent} on the \textit{dev} and \textit{eval} sets, respectively, when initialized randomly. 
\begin{table}[bt]
    \robustify\bfseries
    \sisetup{detect-weight=true, detect-family=true}
    \caption{Comparison of the frame-level diarization with the CHiME7 baseline on DipCo }
    \label{tab:dipco_results}
    \centering
    \begin{tabular}{l S S}
    \toprule
     Model    &  \multicolumn{2}{c}{DER}\\
      & {dev} & {eval} \\
      \midrule
      CHiME7 baseline \cite{23_cornell_chime7}   & 29.9 & 27.9\\
      k-Means & 32.7 & 31.9\\
      VMFMM & \bfseries 24.4 & \bfseries 25.6\\
    $\quad$ + k-Means init & 25.2 & \bfseries 25.6\\
      \bottomrule
    \end{tabular}
\end{table}

\section{Conclusions}
\label{sec:conclusion}
In this work, we proposed a clustering-based method for the diarization of meeting data using frame-level speaker embeddings. It was shown that a simple k-Means clustering is on par with currently used segment-level diarization approaches.
Using k-Means to initialize the components of a von-Mises-Fischer mixture model further stabilized and improved the diarization performance by enhancing the diarization of single-speaker regions, and for a careful parameter choice also of overlapping speech.  
However, the current model still poses a trade-off between a good single-speaker and overlap accuracy. Therefore, we will aim to explicitly allow multiple active classes 
in the future to avoid the thresholding operation on the concentration parameter of the mixture model that currently limits the overlap detection. 

\section{Acknowledgements}
Computational Resources were provided by BMBF/NHR/PC2.
Christoph  Boeddeker was funded by DFG, project no.\ 448568305. 
\pagebreak
\bibliographystyle{IEEEbib}
\bibliography{strings,refs}

\end{document}

%% file: tikz/teacher_student.tex
\begin{tikzpicture}[semithick,auto,
block_high/.style={
		rectangle,
		draw,
		fill=black!20,
		text centered,
		text width=3.5em,
		rounded corners,
 		minimum height=2.5em,
		minimum width=3.5em},
block/.style={
	rectangle,
	draw,
	fill=black!20,
	text centered,
	text width=3.5em,
	rounded corners,
	minimum height=2em,
	minimum width=3.5em},
 block_small/.style={
	rectangle,
	draw,
	fill=black!20,
	text centered,
	text width=2em,
	rounded corners,
	minimum height=2em,
	minimum width=2em},
 block_small_high/.style={
	rectangle,
	draw,
	fill=black!20,
	text centered,
	text width=2em,
	rounded corners,
	minimum height=3em,
	minimum width=2em},
block_teacher/.style={
		rectangle,
		draw,
		fill=black!20,
		text centered,
		text width=3.5em,
		rounded corners,
 		minimum height=2em,
		minimum width=3.5em},
block_teacher_small/.style={
		rectangle,
		draw,
		fill=black!20,
		text centered,
		text width=2em,
		rounded corners,
 		minimum height=2em,
		minimum width=2em},
mul/.style={
        circle,
        draw,
    },		
	]
\tikzset{>=stealth}
\tikzstyle{branch}=[{circle,inner sep=0pt,minimum size=0.3em,fill=black}]

\tikzset{pics/.cd,
	pic switch closer/.style args={#1 times #2}{code={
		\tikzset{x=#1/2,y=#2/2}
		\coordinate (-in) at (1,0);
		\coordinate (-out) at (-1,0);
		
		\draw [line cap=rect] (-1, 0) -- ++(0.1,0) -- ++(20:1.9);
		
	}}
}

    \pgfdeclarelayer{background1}
    \pgfdeclarelayer{background2}
    \pgfdeclarelayer{mid1}
    \pgfdeclarelayer{mid2}
    \pgfdeclarelayer{foreground}
    \pgfsetlayers{background1,background2,mid1,mid2,main,foreground}

    \node[coordinate] (encoder) {};

	\begin{pgfonlayer}{background2}
		\node [block_teacher] (teacherbg) at ($(encoder.east) + (2cm, 1.75cm)$) {ResNet};
		\node [block_teacher_small, right=0.5cm of teacherbg] (poolingbg) {TAP};
            \node [block_teacher, dashed, right=0.5cm of poolingbg] (projectbg) {Project};
	\end{pgfonlayer}
	\begin{pgfonlayer}{background1}
		\node [draw=black, fill=blue!20, fit={(teacherbg) (poolingbg) (projectbg)}] {};
	\end{pgfonlayer}

	\begin{pgfonlayer}{mid2}
		\node [block_teacher, anchor=north west] (teacherfg) at ($(teacherbg.north west) + (-1.5em, -1.5em)$) {ResNet};
		\node [block_teacher_small, right=0.5cm of teacherfg] (poolingfg) {TAP};
              \node [block_teacher, dashed,right=0.5cm of poolingfg] (projectfg) {Project};

	\end{pgfonlayer}
	\begin{pgfonlayer}{mid1}
		\node [draw=black, fill=blue!20, fit={(teacherfg) (poolingfg) (projectfg)}] {};
	\end{pgfonlayer}
	    \tikzstyle{box} = [draw, dashed, inner xsep=1em, inner ysep=1.25em, line width=0.1em, rounded corners=0.3em]
		\node [box, draw=blue,fit={(teacherbg) (poolingbg) (teacherfg) (poolingfg)(projectbg) (projectfg)}, label={[anchor=north east, align=left, yshift=0.4ex]north east:\color{blue}Teacher}] {};
    \node [block, right=1.8cm of encoder] (student) {ResNet};
    \node [block_small_high] (losspit) at ($(projectbg.east) + (4.5em, -0.5em)$) {$\mathcal{L_{\text{Geo}}}$};
    \node[coordinate, right=0.5cm of encoder] (splitteacher) {};
    \node[block, dashed, right=0.4cm of student] (project) {Project};
    \node[block, right=0.4cm of project] (shortpooling) {local TAP};
    \node [coordinate, right=2cm of project] (splitframe) {};
    \node [branch] (splitframe1) at ($(splitframe) + (1.3em, 0.0em)$) {};

	\node [box, draw=purple,fit={(shortpooling) (student) }, label={[anchor=south west, align=left]south west:\color{purple}Student}] {};

   \begin{pgfonlayer}{background1}
   	\draw[->] ($(teacherbg.west) + (-3em, 0)$) -- node[near start, right, xshift=-2.2em] {\footnotesize{$\mathbf{x}_1$}}  ($(teacherbg.west) + (-0em,0)$);
   \end{pgfonlayer}
   \draw[->] ($(teacherfg.west)+ (-3em, 0em)$)  -- node[near start, right, xshift=-2.2em]  {\footnotesize{$\mathbf{x}_2$}} ($(teacherfg.west) + (-0em,0)$);
 \draw[->] ($(encoder.east)$) --node[below, xshift=-0.8em] {\footnotesize{$\mathbf{y}$}} (student);
   \draw[->] ($(projectbg.east) +(0.35em,0)$) -- node[at end, above, xshift=-0.8em,yshift=-0.2em] {\footnotesize{$\mathbf{d}_1$}}($(losspit.west) +(0,0.5em)$);
   \draw[->] ($(projectfg.east) +(0.35em,0)$) -- node[at end,below, xshift=-0.9em, yshift=-0.em] {\footnotesize{$\mathbf{d}_2$}} ($(projectfg.east) +(4.7em,0)$);
   \draw[->] (teacherbg) -- (poolingbg);
  \draw[->] (teacherfg) -- (poolingfg);
  \draw[-] (poolingbg) -- ($(projectbg.west) +(0em,0)$);
 \draw[-] (poolingfg) -- ($(projectfg.west) +(0em,0)$);
  \draw[-] (projectbg) -- ($(projectbg.east) +(0.35em,0)$);
 \draw[-] (projectfg) -- ($(projectfg.east) +(0.35em,0)$);
   \draw[->] ($(student.east)+(0,0em)$) -- ($(project.west)+(0,0em)$);
   \draw[->] (project) -- (shortpooling);

   \draw[-] ($(shortpooling.east)+(0,0em)$) -- node[near end, above, xshift=1.5em, yshift=0em] {\footnotesize{$\mathbf{\hat{d}}(t)$}} ($(splitframe)+(1.3em,-0em)$);
   \draw[->] ($(splitframe1)$) -- ($(losspit.south) +(-0em,0)$);
   \draw[->] ($(splitframe1)$) -- ($(splitframe1.west) +(2em,0.0em)$);

\end{tikzpicture}%

%% file: tikz/concentration_performance.tex
\begin{tikzpicture}
    \begin{axis}[
    xmin=0, xmax=210,
    xtick=data,
    ymin=0, ymax=27.5,
    legend style={at={(1,1)},
          anchor=north east,legend columns=2},
    xlabel ={maximal concentration $\kappa_{\text{max}}$},
    ylabel = {DER [\si{\percent}]},
    xtick pos=left,
    ytick pos=left,
    height=5cm,
    xlabel near ticks,
    ylabel near ticks,
    width=0.5\textwidth,
    ymajorgrids
]
\definecolor{royalblue72120208}{RGB}{72,120,208}
\definecolor{coral23813374}{RGB}{238,133,74}

\addlegendimage{color=black, line width=1pt}
\addlegendentry{avg}
\addlegendimage{color=royalblue72120208,  only marks, mark=*}
\addlegendentry{k-Means}
\addlegendimage{color=black,  dashed, line width=1pt}
\addlegendentry{single}
\addlegendimage{color=coral23813374,  only marks, mark=triangle*}
\addlegendentry{random}
\addplot [royalblue72120208, mark=*, line width=1pt] table {
10   13.18   
25   11.07
50   12.19
75  12.57
100  13.19
150  13.01
200  12.63
};
\addplot [royalblue72120208, mark=*, dashed, line width=1pt, mark options={solid}] table {
10   6.55   
25   6.17
50   4.91
75  4.89
100  5.22
150  4.96
200 4.50
};
\addplot [coral23813374, mark=triangle*, line width=1pt] table {
10   19.76 
25   16.15
50   15.70
75  16.49
100  17.19
150  18.89
200  15.73
};
\addplot [coral23813374,  dashed, mark=triangle*, line width=1pt, mark options={solid}] table {
10  13.84
25  9.10
50  8.17
75 9.07
100 9.36
150 10.97
200 7.86
};
\end{axis}
\end{tikzpicture}